\documentclass[twocolumn,showpacs,preprintnumbers,amsmath,amssymb]{revtex4-1}
\usepackage{epsf}
\usepackage{graphicx}
\usepackage{bm}
\usepackage{psfrag}

\usepackage{pstricks,pst-node,pst-text,pst-3d}
\usepackage{epsf}
\usepackage{epsfig}
\usepackage{subfigure}

\begin{document}

\title{Norm preserving stochastic field equation for an ideal Bose gas in 
a trap:
 \\numerical implementation and applications}

\author{Sigmund Heller}

\author{Walter T. Strunz}%
\affiliation{%
Institut f\"{u}r Theoretische Physik, Technische Universit\"{a}t Dresden, D-01162 Dresden, Germany
}%

\date{\today}

\begin{abstract}
Stochastic field equations represent a powerful tool to describe the thermal 
state of a trapped Bose gas. Often, such approaches are
confronted with the old problem of an ultraviolet catastrophe, which demands 
a cutoff at high energies. In \cite{Hel09} we introduce a quantum stochastic 
field equation, avoiding the cutoff problem through a fully quantum approach
based on the Glauber-Sudarshan $P$-function.
For a close link to actual experimental setups the theory is 
formulated for a fixed particle number and thus
based on the canonical ensemble. In this work the derivation and 
the non-trivial 
numerical implementation of the equation is explained in detail. 
We present applications for finite Bose gases trapped in a variety
of potentials and show results for ground state occupation numbers and
their equilibrium fluctuations. Moreover, we investigate spatial 
coherence properties
by studying correlation functions of various orders.
\end{abstract}

\maketitle
\section{Introduction}
The enormous progress in experimental and theoretical studies of 
ultracold quantum gases leads to a much deeper understanding of quantum 
many body physics \cite{Pet02,Pit03,Blo08}. In particular, Bose-Einstein 
condensation in traps enables us to investigate profoundly quantum statistical
phenomena in finite systems. In more recent experiments,
spatial and temporal coherences \cite{Blo00,Hel03,Foe05,Sch05,Oet05} 
of ultracold Bose gases are investigated in terms of correlation
functions. Moreover, dynamics from 
nonequilibrium to equilibrium states \cite{Rit07} and the spatial
dependence of equilibrium density 
fluctuations \cite{Est06} are considered. 
For a theoretical description of these phenomena it is desirable to 
consider nonzero temperature, finite size and -- as no particle bath
is present -- a canonical description of the many-body quantum system.
For most experiments, interaction between the atoms are of crucial 
importance. In systems with a Feshbach resonance, the interaction
strength may even be tuned over a wide range -- allowing to study 
ideal gases, too. In this article we deal almost exclusively with an ideal gas and 
comment on the interacting case briefly at the end of our paper.

It is clear that many properties of ultracold gases are sensitive to
temperature not least due to the different size of the condensed 
fraction of the gas. 
Moreover, for these finite systems with a fixed number of particles, 
it is preferable to base a theoretical description on the canonical ensemble
rather than the usual grand canonical ensemble. For an ideal gas,
the best known example for
the significance to choose the canonical ensemble is
the fluctuation of the ground state occupation number 
$\langle N^2_0-\langle N_0\rangle^2\rangle$ \cite{Kac77}.
Clearly, it approaches zero as temperature tends to zero in a
canonical ensemble. Yet it is of the size of the average particle number $N$ 
in a grand canonical description. A detailed study of the fluctuations of the 
ground state particle number in the microcanonical and the canonical 
ensembles are given in \cite{Gro96,Pol96,Wil97,Gro97,Hol98,Koc00}. 
Spatial correlations are
among other quantities of relevance for which a
grand canonical formulation differs significantly from a canonical
description \cite{Nar99}. 

Recently, we presented a norm preserving stochastic field equation that
describes
the canonical state of an ideal Bose gas \cite{Hel09}.
The resulting equation fully
reflects quantum statistics and was given in \cite{Hel09} 
in a representation independent form. Therefore, it can
be propagated in position space, meaning that it is not necessary to know 
eigenfunctions and -energies of the trapped atoms. 
While in \cite{Hel09} we simply
state the equation and display a few applications, we here want to
elaborate on its derivation, its numerical implementation and further
applications in much more detail. 

As we will show,
spatial correlations and fluctuations can easily be calculated in position
space. Due to the quantum framework, the equation -- despite representing
a c-number field -- does not suffer from 
any cutoff problems which usually occur for classical field equations. 
Thus, numerical simulations can be performed with sufficient precision 
and reasonable effort. 

Let us relate our approach to previous stochastic equations for the grand 
canonical ensemble. Note that most of these investigations are concerned
with an interacting gas, while here, as a first step, we restrict ourselves
to the case of the ideal gas. Since our theory can be formulated in 
position space, we clearly expect to be able to include
interactions in a mean field sense in a next step. 

Detailed discussions of different approaches to (interacting)
ultracold gases in terms of stochastic field equations are given in \cite{Bra05,Pro08}.
Exact methods based on the positive P-representation of the full density 
operator are used by Drummond and co-workers \cite{Dru04}. This
theory brings about numerical challenges and is applied mainly to 
one-dimensional systems. More recently, also a 3D free gas at temperatures 
well below the transition temperature has been discussed \cite{Dru07}.
Other approaches \cite{Sto97,Sto01,Dav01,Gar02,Bra05} can be seen as 
classical field methods, in which the lowly occupied energy levels must 
be treated in a different formalism or neglected, due to ultraviolet problems. 
Davis and co-workers apply a projection on a subspace of highly occupied 
energy levels, using a cutoff \cite{Dav01}, while Gardiner and co-workers 
separate the field operator in a highly and lowly occupied 
part \cite{Gar02,Bra05}.
We see our approach more in the spirit of Stoof and co-worker, who start from 
a path integral approach and arrive at a functional Fokker-Planck equation 
for the Wigner functional of the thermal field \cite{Sto97}. After additional
simplifications, in their work the corresponding Langevin equation 
takes the form of a classical stochastic 
field equation as given by Hohenberg and Halperin \cite{Hoh77}. 

Apart from these grand canonical approaches,
theories for a fixed particle number have also been presented more recently 
\cite{Xio02,Tri08,Tri082,Bar08}.  Xu and co-workers describe Bose gases 
with first order perturbation theory and in a Bogoliubov approximation 
\cite{Xio02}. An exact phase-space description for a finite number of 
particles is worked out by Korsch and co-workers. Exact evolution equations 
for the Husimi-Q- and the Glauber-Sudarshan-P-distribution function based 
on a Bose-Hubbard Hamiltonian are derived \cite{Tri08,Tri082}. Stochastic 
methods in phase-space for the canonical ensemble of qubit and spin models 
have been presented by Drummond and co-workers \cite{Bar08}. While all
these theories based on a canonical ensemble focus on different systems, 
e.g. lattices or spin models, we here aim to describe the full canonical
thermal state of a Bose gas in any given trap.

Our novel equation presented in \cite{Hel09} has been found on the
basis of the Glauber-Sudarshan-P-representation -- being exact for an
ideal gas. In this paper we aim
to explain in more detail the derivation of the equation, its properties,
its numerical implementation, and we show its power by discussing further
applications.
The paper is structured as follows:
In the second section we show the theoretical background and the 
derivation of the stochastic field equation. The third section is devoted to
the discussion of some crucial properties of the equation. 
In the fourth section we 
demonstrate how we propagate the equation in energy and in position space
numerically. 
Results of these numerical applications are shown in the fifth section, 
before we draw conclusions in the final section.     

\section{Stochastic field equation for the canonical ensemble}
In order to derive a quantum stochastic field equation for the canonical 
ensemble of an ideal Bose gas, our first aim is to express quantum expectation 
values for a fixed particle number in terms of c-number integrals. 
The desired stochastic field equation should enable us to calculate 
these expectation values. We use the corresponding 
Fokker-Planck equation to demonstrate the equivalence of 
stochastic and canonical quantum ensemble mean.

We begin with the canonical density operator for $N$ particles
\begin{equation}\label{density}
\hat{\rho}_N=\frac{1}{Z_N}e^{-\beta\hat{H}}P_N.
\end{equation}
As usual, the Hamiltonian 
$\hat{H}=\sum\limits_k\epsilon_k\hat{a}^{\dagger}_k\hat{a}_k$ 
can be expressed in terms of occupation numbers of single-particle states
($\epsilon_k$ denotes the energy of single-particle state $k$). Furthermore,
the canonical partition function is denoted by $Z_N$, 
with the usual $\beta=\frac{1}{kT}$ incorporating
Boltzmann's constant $k$ and temperature $T$. Crucially,
a projector 
$P_N=\sum\limits_{\sum n_k=N}|\{n_k\}\rangle
\langle\{ n_k \}|$ on the $N$-particle subspace appears, with
the usual number states $|\{n_k\}\rangle=
\frac{1}{\sqrt{\prod\limits_kn_k!}}
\prod\limits_k(\hat{a}^{\dagger}_k)^{n_k}|0\rangle$. 
The operator $P_N$ projects the exponential $e^{-\beta\hat{H}}$ on a 
subspace of $N$ particles. In order to arrive at a c-number representation of 
quantum expectation values, one can express $e^{-\beta\hat{H}}$ 
from (\ref{density}) as a mixture of coherent states using the 
(Glauber-Sudarshan) P-function representation \cite{Sch01}
\begin{eqnarray}\label{p-function}
\frac{e^{-\beta\hat{H}}}{Z}&=&\frac{1}{\bar{n}} \int d\mu\{z\}
\exp\left(-\sum\limits_k|z_k|^2(e^{\beta \epsilon_k}-1)\right)
\nonumber\\&&\,\times|\{z\}\rangle\langle\{z\}|\nonumber\\&=&:
\int d\mu\{z\}P(\{z\})|\{z\}\rangle\langle\{z\}|.
\end{eqnarray}
Here, we abbreviate the normalization factor with 
$\bar{n}=\prod\limits_k(e^{\beta{\epsilon_k}}-1)^{-1}$, and products of coherent 
states with $|\{z\}\rangle= |z_0\rangle|z_1\rangle\cdots|z_n\rangle\cdots$.
The appropriate measure is 
$d\mu\{z\}=\prod\limits_k\left(\frac{d\{\text{Re}z_k\}d\{\text{Im}z_k\}}
{\pi}\right)$. 

Of particular interest are
correlation functions of quantum field operators. Using the P-representation 
(\ref{p-function}), we can determine canonical quantum expectation values
with the help of
$\langle \{z\} |P_N|\{z\}\rangle=\frac{1}{N!}(\sum\limits_k|z_k|^2)^N 
e^{-\sum\limits_k|z_k|^2}$ (see \cite{Mol67}) and obtain
a weight function 
$W_N(\{z\})=\frac{1}{N!}\left(\sum\limits_{k}|z_k|^2\right)^{N}
e^{\left(-\sum\limits_k|z_k|^2\right)}P(\{z\})$. With the latter,
quantum expectation values can be written as integrals over c-numbers,
\begin{eqnarray}
\langle \hat{a}^{\dagger}_i\hat{a}_j\rangle_N=\text{tr}\left(
\hat{a}^{\dagger}_i\hat{a}_j\,\hat\rho_N\right)
=
\frac{1}{C}\int d\mu\{z\}z^{\ast}_i z_j W_{N-1}(\{z\})\nonumber\\\label{ensembleaverage}
\end{eqnarray} 
with $C=\int d\mu\{z\}W_{N}(\{z\})$.

In this work we focus on spatial correlation functions. Therefore, we are
interested in ensemble averages for the bosonic field operator
$\hat{\psi}(x)=\sum\limits_k\langle x|\epsilon_k\rangle\hat{a}_k$
with $|\epsilon_k\rangle$ the single-particle eigenstate corresponding
to energy $\epsilon_k$. First, the weight function is 
rewritten as $W_N(\{\psi\})=\frac{1}{N!}
\left(\int dx\,|\psi(x)|^2\right)^{N}e^{\left(-\int dx\,|\psi(x)|^2\right)}
P(\psi^{\ast},\psi)$ using $\psi(x)=\sum\limits_kz_k\langle x
|\epsilon_k\rangle$ which is the eigenvalue of the field operator 
when applied to the coherent state $|\{z\}\rangle$, 
i.e. $\hat{\psi}(x)|\{z\}\rangle=\psi(x)|\{z\}\rangle$. 
Further, we obtain  
\begin{eqnarray}
\langle \hat{\psi}^{\dagger}(x)\hat{\psi}(x')\rangle_N &=&
\frac{1}{C}\int\, d\mu\{ \psi\}\psi^{\ast}(x)\psi(x')W_{N-1}(\{\psi\})\nonumber\\
\end{eqnarray} 
with the normalisation constant $C=\int d\mu\{\psi\}W_{N}(\{\psi\})$. 
It should be mentioned here that for quantum expectation values of second 
(or higher) order one has to use weight functions of different $N$. 
For $\langle\hat{\psi}^{\dagger}(x)\hat{\psi}^{\dagger}(x')
\hat{\psi}(x)\hat{\psi}(x')\rangle$, for instance, the first order 
$W_{N-1}(\{\psi\})$ has to be replaced by
$W_{N-2}(\{\psi\})$ --
for higher orders the result changes accordingly (k-th order: 
chose $W_{N-k}(\{\psi\})$). 

In order to calculate these expectation values we were able to
construct a new, norm-preserving stochastic field equation \cite{Hel09}.
The equivalence of quantum statistical and stochastic ensemble mean
is based on the fact that the weight functions $W_n$ ($n=N, N-1, \ldots$)
turn out to be stationary solutions of the corresponding 
Fokker-Planck equation.

In representation independent form ($\psi(x,t)=\langle x|\psi(t)\rangle$) 
and using Stratonovich calculus \cite{Gar83}
it reads ($\hbar=1$ throughout)
\begin{eqnarray}\label{quantumequation}
\text {(S)  }d|\psi\rangle=-\left((\Lambda+i)H-\Lambda\frac{\langle\psi|H|\psi\rangle}{\langle\psi|K|\psi\rangle}K\right)|\psi\rangle dt\nonumber\\+\sqrt{2\Lambda}\left(\sqrt{K}|d\xi\rangle-\frac{\langle\psi|\sqrt{K}|d\xi\rangle}{\langle\psi|K|\psi\rangle}K|\psi\rangle\right).
\end{eqnarray}
Here, the single particle Hamiltonian $H=\frac{p^2}{2m}+V(x)
=\sum\limits_k\epsilon_k|\epsilon_k\rangle\langle\epsilon_k|$ appears, 
and a damping parameter $\Lambda$. Crucially, temperature enters through
an {\it operator}
\begin{equation}\label{Koperator}
K=\frac{H}{e^{\beta H}-1}.
\end{equation}
The noise increment $|d\xi\rangle$ is 
uncorrelated in space and time 
$\langle x| d\xi(t)\rangle\langle d\xi(t)| x' \rangle=\delta(x-x')dt$. 
Equation (\ref{quantumequation}) is a
norm preserving, non-linear stochastic equation that enables us to 
obtain canonical expectation values on average.

It is important to stress that the dependence of
the stochastic fields $|\psi(t)\rangle$ on
the particle number $N$ is indirectly given through their norm 
${\cal N}= \langle\psi|\psi\rangle$ which is preserved during
propagation and thus determined by the initial condition.
One might expect ${\cal N}\approx N$ to be a reasonable choice;
however, matters are more delicate and a distribution of norms
${\cal N}$ is required in order to obtain exact quantum
expectation values. We devote the whole next Section \ref{norm}
to the relation between norm ${\cal N}$ and particle number $N$.

Sometimes it useful to express (\ref{quantumequation}) as an 
Ito-stochastic equation, reading
\begin{eqnarray}\label{quantumequationIto}
\text {(I)  }d|\psi\rangle=
-\left((\Lambda+i)H-\Lambda\frac{\langle\psi|H|\psi\rangle}
{\langle\psi|K|\psi\rangle}K\right)|\psi\rangle dt\nonumber\\
-\frac{\Lambda}{\langle\psi|K|\psi\rangle}
\left(\text{tr}(K)-\frac{\langle\psi|K^2|\psi\rangle}
{\langle\psi|K|\psi\rangle}\right)K|\psi\rangle dt\nonumber\\
+\sqrt{2\Lambda}\left(\sqrt{K}|d\xi\rangle
-\frac{\langle\psi|\sqrt{K}|d\xi\rangle}
{\langle\psi|K|\psi\rangle}K|\psi\rangle\right).
\end{eqnarray}

\begin{figure*}[h,t]
  \centering
  \fbox{
    \includegraphics[width=8.3cm]{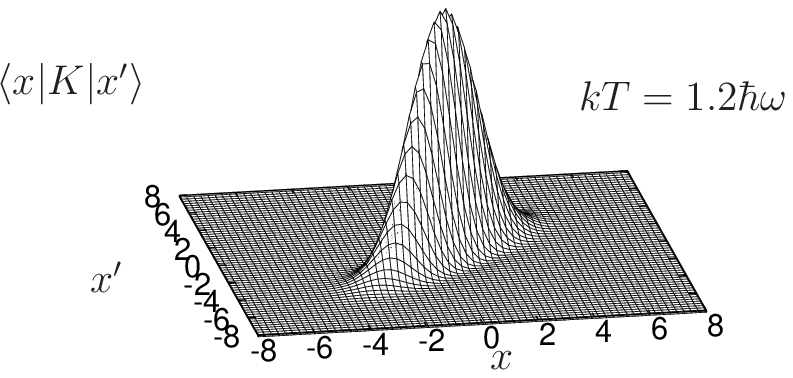}
\includegraphics[width=8.3cm]{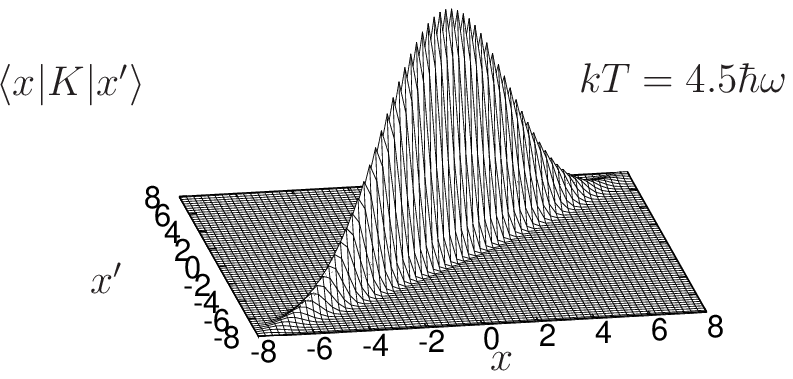}
  }
  \caption{Correlation function $\langle x|K|x'\rangle$ of the noise
$|d\zeta\rangle = \sqrt{K}|d\xi\rangle$ of our stochastic field 
equation  (\ref{quantumequation}) for a harmonic potential.
We show the correlation function for temperature $kT=1.2 \hbar \omega$ (left)
and for a higher temperature $kT=4.5 \hbar \omega$ (right).}\label{b1}
 \end{figure*}

The novel stochastic field equation can be easily solved numerically and used in different 
representations. An implementation in the position representation 
is particularly useful as it can be applied to arbitrary external 
potentials $V(x)$. Based on the assumption of sufficient ergodicity, in applications the ensemble 
averages are replaced by a long time limit 
$\langle\hat{\psi}^{\dagger}(x)\hat{\psi}(x')\rangle_N=
\lim\limits_{t\rightarrow\infty}\frac{1}{t}\int\limits_0^{t}ds\,
\{\psi^{\ast}(x,s)\psi(x',s)\}$ over a single realization $\psi(x,s)$ 
of the stochastic field equation. 

The term $(\Lambda+i)$ consists of a phenomenological damping term 
$\Lambda$, while "$i$" describes "real" dynamics. So, 
in a phenomenological manner, our equation
can also be used to mimic the transition from a non-equilibrium 
to an equilibrium state.

It is important to point out the crucial role of the operator $K$. 
Its meaning becomes most apparent when
we omit the nonlinear terms in our equation (\ref{quantumequation}) 
and think of $H$ to represent $H-\mu$ which also
affects the operator $K$.
One obtains an equation 
which gives us the thermal state of the grand canonical ensemble \cite{Hel07}
\begin{eqnarray}
 d|\psi\rangle=-(\Lambda+i)H|\psi\rangle dt +\sqrt{2\Lambda K}|d\xi\rangle.\label{quantumgc}
\end{eqnarray}
with $\mu$ the chemical potential used to fix the average particle number $N$.
It should be mentioned here that the canonical equation 
(\ref{quantumequation}) is not merely a 
normalized version of the grand canonical equation. The linear equation 
(\ref{quantumgc}) is closely related to the classical field equation 
for the thermal state of the grand canonical ensemble for 
any Hamiltonian energy functional ${\cal H}(\psi,\psi^*)$ \cite{Hoh77} 
\begin{eqnarray}
 d\psi(x,t)=-(\Lambda+i)\frac{\delta {\cal H}}{\delta \psi^*} dt
+\sqrt{2\Lambda kT}d\xi(x,t),\label{classicalgc}
\end{eqnarray}
which has well-known ultraviolet problems.
If one sets $\frac{\delta {\cal H}}{\delta \psi^*} =H\psi$, there is only one
difference between classical (\ref{classicalgc}) and quantum equation
(\ref{quantumgc}): the classical 
temperature $kT$ is substituted by the operator $K$. As discussed
in \cite{Dav01}, for instance, the classical equation satisfies the 
equipartition theorem and therefore the infinite degrees of freedom 
of a field $\psi(x,t)$ lead to an ultraviolet 
catastrophe. This can also be seen in equ. (\ref{classicalgc}) looking at 
a formulation in position space: the white noise term which is uncorrelated
in position induces arbitrarily high momentum kicks. However, if we consider the operator $K=\frac{H}{e^{\beta H}-1}$ acting on the white noise $|d\xi\rangle$ in position space representation, the former white noise becomes 
spatially correlated, 
and we no longer get unphysical momentum. In other words,
proper quantum statistics requires the replacement of spatially uncorrelated 
noise $\sqrt{kT}|d\xi\rangle$ in (\ref{classicalgc}) by spatially correlated effective noise 
$|d\zeta\rangle = \sqrt{K}|d\xi\rangle$, as in (\ref{quantumgc}). The correlation 
function of the effective noise
$\langle x| d\zeta(t)\rangle\langle d\zeta(t)|x'\rangle 
= \langle x|K| x' \rangle\,dt$ is now a smooth function in 
position space and shown in Fig. \ref{b1} for two different temperatures.

In the low energy limit $H\ll kT$, indeed, the operator 
tends to the classical temperature $K\approx kT$. Yet in the case of 
high energies $H\gg kT$, the operator tends to zero $K\rightarrow 0$. 
We see that in the quantum equations (\ref{quantumequation}, \ref{quantumgc}) 
the required cutoff is built in. 

\section{Stochastic average and quantum expectation value\label{norm}}
In this chapter we want to investigate the relationship between 
quantum expectation values of arbitrary order 
$\langle \hat{a}^{\dagger}_i...\hat{a}^{\dagger}_j\hat{a}_k...\hat{a}_l\rangle$
(\ref{ensembleaverage}) and those obtained with the stochastic field 
equation (SFE) 
$\langle\langle z^{\ast}_i...z^{\ast}_jz_k...z_l\rangle\rangle_{\text{SFE}}$. 
For the following we introduce the norm
${\cal N}\equiv\sum\limits_k|z_k|^2=\langle\psi|\psi\rangle$ of the 
wavefunction. The stochastic equation keeps the norm fixed, while the 
integral of the quantum expectation values (\ref{ensembleaverage}) 
extends over a range of values of the quantity
$\sum\limits_k|z_k|^2$. The norm stays constant and for any
stochastic method $\langle\langle 1 \rangle\rangle_{\text{SFE}}=1$.
Hence, we can write the first order expectation values we obtain from the 
stochastic equation as
\begin{eqnarray}
 &&\langle\langle z^{\ast}_iz_j\rangle\rangle_{\text{SFE}}({\cal N},
\epsilon_0)\nonumber \\&&\equiv
\frac{\int d\mu\{z\}z^{\ast}_iz_j\delta({\cal N}-\sum\limits_k|z_k|^2)
W_{N-1}(\{z\})}
{\int d\mu\{z\}\delta({\cal N}-\sum\limits_k|z_k|^2)W_{N-1}(\{z\})}.
\end{eqnarray}
We find that this expression depends on the ground state energy 
$\epsilon_0$. It can be seen easily by a substitution 
$z\rightarrow z e^{\beta \epsilon_0/2}$ that the simple identity
$e^{\beta \epsilon_0}\langle\langle z^{\ast}_iz_j\rangle\rangle_{\text{SFE}}
({\cal N}e^{-\beta \epsilon_0},\epsilon_0)=
\langle\langle z^{\ast}_iz_j\rangle\rangle_{\text{SFE}}({\cal N},0)$ holds.
The relation between the average values given by the stochastic field 
equation and quantum expectation values is now obvious
\begin{eqnarray}
\label{SFEqm}
 \langle\hat {a}^{\dagger}_i\hat{a}_j\rangle_N&=&
\int d{\cal N} {\tilde{P}} ({\cal N}) 
\langle\langle z^{\ast}_iz_j\rangle\rangle_{\text{SFE}}({\cal N},\epsilon_0).
\end{eqnarray}
with the norm distribution
\begin{eqnarray}
{\tilde{P}}({\cal N})\equiv\frac{\int d\mu\{z\}\delta({\cal N}-\sum\limits_k|z_k|^2)W_{N-1}(\{z\})}{\int d\mu\{z\}W_{N}(\{z\})}
\label{weightterm}    
\end{eqnarray}
in the integral (\ref{SFEqm}).\\ 

Note that $\int d{\cal N} {\tilde{P}} ({\cal N}) =
\int d\mu\{z\}W_{N-1}(\{z\})/\int d\mu\{z\}W_{N}(\{z\})\neq 1$. Therefore,
and in order to get a better understanding of equ. (\ref{SFEqm}), we
have to investigate ${\tilde{P}} ({\cal N})$ more closely. A
lengthy calculation is given in appendix \ref{A1}, which results in

\begin{eqnarray}\label{pdistribution}
{\tilde{P}}({\cal N})
=\frac{1}{(N-1)!}\frac{{\cal N}^{(N-1)}
\sum\limits_{k}c_ke^{-e^{\beta \epsilon_k} {\cal N}}}
{\sum\limits_kc_ke^{-\beta \epsilon_k(N+1)}}.\nonumber\\ 
\end{eqnarray}
with $c_k=\prod\limits_{\begin{array}{c}\scriptstyle l=0\scriptstyle \\ 
\scriptstyle l\neq k
\end{array}}\left(e^{\beta \epsilon_l}-e^{\beta \epsilon_k}\right)^{-1}$.
Well below the critical temperature it is sufficient to consider only the 
first term of the sum $\sum\limits_{k}c_ke^{-e^{\beta \epsilon_k} {\cal N}}$ 
because the remaining terms are smaller by a factor 
$e^{-(e^{\beta \epsilon_k}-e^{\beta \epsilon_0}) {\cal N}}$. 
Numerical investigations of the factors $c_k$ showed that this 
approximation is justified in the temperature region considered here. 
Then our distribution ${\tilde{P}}({\cal N})$ is normalized to   
\begin{eqnarray}
&&\int\limits_{0}^{\infty}d{\cal N} {\tilde{P}}({\cal N})=
\frac{\sum\limits_k c_k e^{-\beta \epsilon_kN}}
{\sum\limits_k c_k e^{-\beta \epsilon_k(N+1)}}
\approx e^{\beta \epsilon_0}.\label{averagenorm}
\end{eqnarray}
Within this approximation we have a Poisson-like distribution 
\begin{equation}
{\tilde{P}}({\cal N}) \sim
{\cal N}^{(N-1)}\exp(-e^{\beta \epsilon_0}{\cal N}).
\label{dis-approx}
\end{equation}
This distribution is centered around 
\begin{eqnarray}
\langle{\cal N}\rangle=
\frac{\int\limits_{0}^{\infty}d{\cal N} 
{\tilde{P}}({\cal N}){\cal N}}
{\int\limits_{0}^{\infty}d{\cal N} {\tilde{P}}({\cal N})}
=Ne^{-\beta \epsilon_0}\label{normnumber}
\end{eqnarray}
and the standard deviation equals 
$\Delta {\cal N}=\sqrt{N} e^{-\beta \epsilon_0}$. For large $N$ the 
ratio of variance and mean goes to zero 
$\frac{\Delta {\cal N}}{\langle {\cal N}\rangle}\rightarrow 0$. Hence, 
in general it is sufficient to propagate equ. (\ref{quantumequation})
with a single norm, i.e. we replace
${\tilde{P}}({\cal N})\approx e^{\beta \epsilon_0}
\delta({\cal N}-Ne^{-\beta \epsilon_0})$. 
As a consequence, from (\ref{SFEqm}) we get the simple relation 
$\langle\hat {a}^{\dagger}_i\hat{a}_j\rangle_N
=e^{\beta \epsilon_0}\langle\langle z^{\ast}_iz_j\rangle\rangle_{\text{SFE}}
(Ne^{-\beta \epsilon_0},\epsilon_0)$ between canonical quantum
correlation function and ensemble average of the stochastic field equation. 
As further elaborated upon in Section \ref{Results}, it is
for very high precision only, that
it is necessary to take into account the full norm distribution.
Let us also remark that for higher order expectation values these 
considerations apply similarly.

It should be noted here that if the norm of our numerical simulation of the 
stochastic field equation ${\cal N}$ should correspond to the particle 
number $N$, we have to choose $\epsilon_0=0$. However, it is also possible 
to choose a norm different from the particle number according to
(\ref{normnumber}). This freedom is useful for an easier numerical 
implementation, as shown in Section \ref{ni}.
It should also be mentioned that for temperatures above and near the 
critical temperature, the approximations leading to expression
(\ref{averagenorm}) are no longer valid and the exact norm distribution
needs to be known. Still, using (\ref{averagenorm}), for many quantities 
we observe satisfying results even in this temperature region.

\section{Numerical implementation \label{ni}}
In this section it is shown how one can solve the novel stochastic field 
equation numerically. The implementations can be done in different 
representations, whichever is convenient for a given trap potential. 
If the single particle eigenenergies are known (recall that we neglect
interatomic interactions here) a propagation in energy space is the easiest 
way to solve our equation. In a box potential this is analogous to a 
propagation in momentum space. For a trap potential $V(x)$ with unknown energy 
spectrum, one can solve the equation in position space directly, using 
a split operator method based on Fast Fourier Transformation. This 
implementation is more challenging but the program can be easily adjusted to any trap just by changing the single line of code where the potential 
is defined. 

\begin{figure*}[h,t]
  \centering
  \fbox{
    \includegraphics[width=8.3cm]{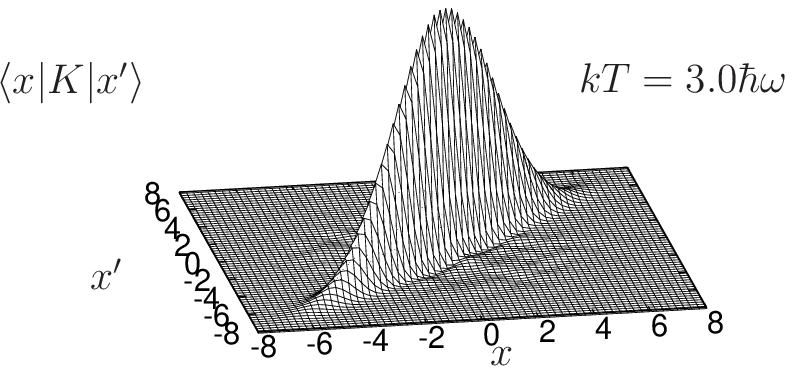}
\includegraphics[width=8.3cm]{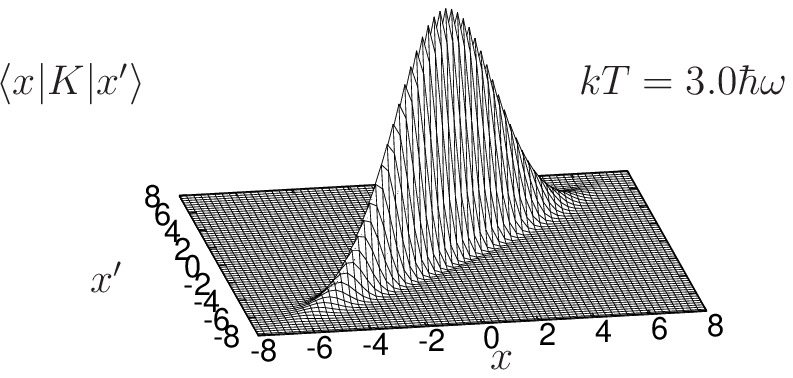}
  }
  \caption{Correlation function $\langle x|K|x'\rangle$ of the correlated 
noise of our stochastic field equation (\ref{quantumequation}) for a 
harmonic potential at temperature $kT=3 \hbar \omega$.
We show the correlation function obtained from the noise in Wigner-Weyl
approximation (left) compared to the exact result (right).}\label{b2}
 \end{figure*}


If the eigenvalues for the given trap potential are known, a numerical 
implementation in energy space does not contain any challenging aspects. 
As already explained above, it is more convenient to pass to position space, 
if we want to solve problems with arbitrary external potential, 
for which the eigenenergies are not known. 
We turn to a discretized formulation of equation (\ref{quantumequationIto})
and use a position grid $\{x_i\}$ with some multi-index $i$.
The part of the equation containing the single-particle
Hamiltonian $H=\frac{p^2}{2m}+V(x)$ only can be propagated easily using 
standard Fast-Fourier transformation methods \cite{Fei82}. 

The challenging aspect is the evaluation of
matrix elements of the operator (\ref{Koperator}) appearing in
(\ref{quantumequationIto}), i.e. the determination of
$\langle x_i|K|x_j\rangle=
\left\langle x_i\left|\frac{\frac{p^2}{2m}+V(x)}
{e^{\beta\left(\frac{p^2}{2m}+V(x)\right)}-1}\right|x_j\right\rangle$.

For an efficient implementation a Wigner-Weyl-approximation is 
helpful. The Wigner-Weyl correspondence \cite{Sch01} of the operator 
$K$ in $2D$-dimensional phase space is defined as 
\begin{eqnarray}
K(x,p)=\int d^Ds\, e^{ips}\langle
x+\frac{s}{2}|K|x-\frac{s}{2}
\rangle,
\end{eqnarray}
with the help of which we can obtain the matrix elements as 
\begin{eqnarray}
\langle x_i|K|x_j\rangle=\frac{1}{(2\pi)^D}\int d^Dp\, e^{ip(x_i-x_j)}
K\left(\frac{x_i+x_j}{2},p\right),\nonumber\\
\end{eqnarray}  
In order to simplify, we consider only zeroth order in $\hbar$, and find the classical 
expression
\begin{eqnarray}
K(x,p)\approx \frac{\left(V(x)+\frac{p^2}{2}\right)}
{e^{\beta\left(V(x)+\frac{p^2}{2}\right)}-1},
\end{eqnarray}
This approximation will be justified by comparing the approximated 
numerical result with exact numerical calculations.

Expanding the denominator in a geometric series we get 
\begin{eqnarray}
\langle x_i|K|x_j
\rangle&\approx&\frac{1}{(2\pi)^D}\int d^Dp
e^{ip(x_i-x_j)}\left(V\left(\frac{x_i+x_j}{2}\right)\right.\nonumber\\&&
\left.\vphantom{\sum_{i=1}^{\lceil\frac{n}{2}\rceil}}+\frac{p^2}{2}\right)\sum\limits_{n=1}^{\infty}e^{-\beta
  n \left(V\left(\frac{x_i+x_j}{2}\right)+\frac{p^2}{2}\right)}\label{beginnoise}
\end{eqnarray}

\begin{figure*}[h,t]
  \centering
  \fbox{
    \includegraphics[width=8.4cm]{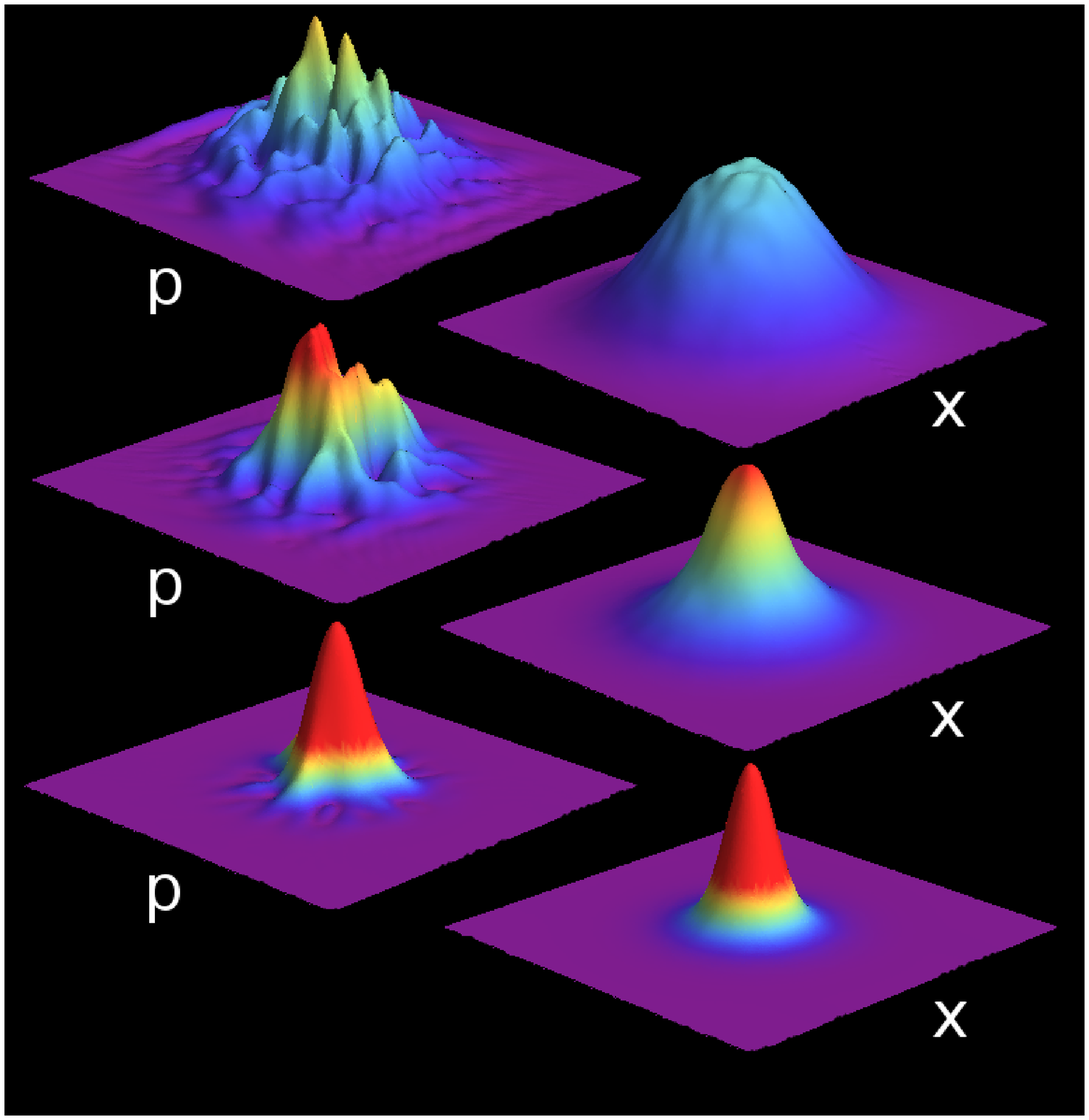}
\includegraphics[width=8.2cm]{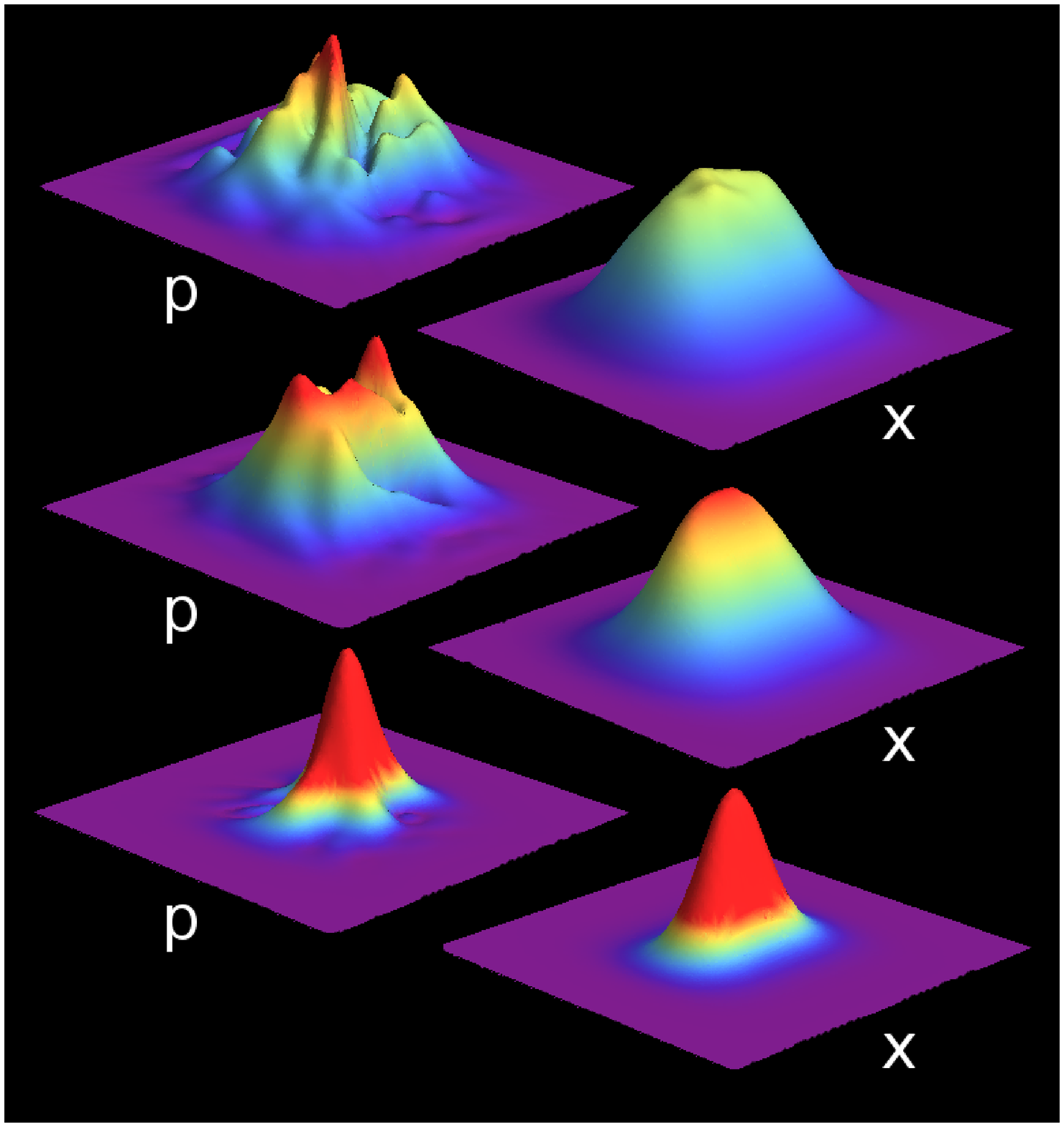}
  }
  \caption{Wigner function of a Bose gas of $100$ particles trapped
in a harmonic potential (left) and in the quartic potential 
$V(x,y,z)=x^4+y^4+z^4$ (right) above (top), at (middle), and below (bottom) 
the critical temperature. On the left hand side of both pictures we display 
a single realization with our stochastic field equation 
(\ref{quantumequation}), on the right hand side a long-time average over 
15000 time steps.\\
  Top: almost a Boltzmann distribution for $T>T_c(N)$;
  middle: a peak develops for $T\approx T_c(N)$;
  bottom: almost all particles condense in the ground state, the Wigner 
  function reflects the ground state Wigner function $T<T_c(N)$.}
  \label{b3}
 \end{figure*}

The momentum-integration can now be carried out without any difficulty 
which leads to the following expression:
\begin{eqnarray}
\langle x_i|K|x_j\rangle&\approx&\frac{1}{(2\pi\beta)^{\frac{D}{2}}}
\sum\limits_{n=1}^{\infty}e^{-\beta n V\left(\frac{x_i+x_j}{2}\right)}
\left(V\left(\frac{x_i+x_j}{2}\right)
\right.\nonumber\\&&
\left.+\frac{D}{2\beta n}-\frac{(x_i-x_j)^2}{2(\beta n)^2}\right)
e^{\frac{-(x_i-x_j)^2}{2\beta n}}.\label{correlation}
\end{eqnarray}     

It is important to note that the convergence of the geometric series
in (\ref{beginnoise}) very much depends on the minimum of the potential
energy. Indeed, rapid convergence can be assured by adding some
(physically irrelevant) constant to $V(x)$. However, as this shift
also affects the ground state energy $\epsilon_0$, it also has a 
significant influence on the
norm distribution of the wave function as can be witnessed 
in (\ref{normnumber}).
Therefore, in practice one tries to identify some optimal shift
that assures both, rapid convergence and a reasonable norm of the
wave function.

An efficient generation of $\langle x_i|K|x_j\rangle$
is given in equ. (\ref{correlation}). Now, in order to propagate equ. 
(\ref{quantumequationIto}), it is necessary to think about the numerical 
implementation of the correlated noise $|d\zeta\rangle = \sqrt{K}|d\xi\rangle$,
too.
Obviously, the (discretized) noise $d\zeta_i=\langle x_i|d\zeta\rangle$ 
must be generated in a way that 
\begin{equation}
d\zeta_i d\zeta_j^* = \langle x_i |K|x_j\rangle dt
\nonumber\\
\end{equation}
holds. In order to find an efficient way to obtain the noise, we start 
with equ. (\ref{beginnoise}).  
Based on the substitution $p\rightarrow\sqrt{n\beta}p$ we get
\begin{eqnarray}
\langle x_i|K|x_j\rangle&\approx&\frac{1}{(2\pi\sqrt{n\beta})^D}
\sum\limits_{n=1}^{\infty}e^{-\beta
  n V\left(\frac{x_i+x_j}{2}\right)}\nonumber\\&&
\left(V\left(\frac{x_i+x_j}{2}\right)\int d^Dp e^{i\frac{p(x_i-x_j)}{\sqrt{n\beta}}}e^{-\frac{p^2}{2}}\right.\nonumber\\&&
\left.\vphantom{\sum_{i=1}^{\lceil\frac{n}{2}\rceil}}+\frac{1}{n\beta}\int d^Dp \frac{p^2}{2}e^{i\frac{p(x_i-x_j)}{\sqrt{n\beta}}}e^{-\frac{p^2}{2}}\right).\label{substitution}
\end{eqnarray}
For simplicity, the following derivation of an efficient noise generating
method will be restricted to the one-dimensional case $D=1$. For
higher dimensions, the strategy follows the very same lines.\\
The first integral $\int dp e^{i\frac{p(x_i-x_j)}{\sqrt{n\beta}}}
e^{-\frac{p^2}{2}}$ can be obtained from a Monte-Carlo integration
of $\langle\!\langle e^{i\frac{\eta_1(x_i-x_j)}{\sqrt{n\beta}}}
\rangle\!\rangle$ over a normally distributed real random variable $\eta_1$.
For the second momentum integral in (\ref{substitution}), note that
the integral $\int dp\frac{p^2}{2}e^{i\frac{p(x_i-x_j)}{\sqrt{n\beta}}}
e^{-\frac{p^2}{2}}$ can be seen as
a radial integration in spherical coordinates according to
\begin{eqnarray} 
\int dp \frac{p^2}{2}e^{i\frac{p(x_i-x_j)}{\sqrt{n\beta}}}e^{-\frac{p^2}{2}}
=\int\limits_{0}^{\infty} dp\,
p^2\cos\left(\frac{p(x_i-x_j)}{\sqrt{n\beta}}\right)e^{-\frac{p^2}{2}}
\nonumber\\=\frac{1}{4\pi}\int d^3{\bf p}\;
\cos\left(\frac{\sqrt{p_1^2+p_2^2+p_3^2}(x_i-x_j)}{\sqrt{n\beta}}\right)
e^{-\frac{p_1^2+p_2^2+p_3^2}{2}}.\nonumber\\\label{spherical}
\end{eqnarray}
Obviously, now the integral in equ. (\ref{spherical}) can be evaluated by 
a Monte-Carlo integration using three additional real Gaussian random 
numbers $\eta_2, \eta_3,$ and $\eta_4$. We arrive at the following
representation of the correlation function as an average over four
normally distributed real random variables $\eta_1,\ldots,\eta_4$:
\begin{eqnarray}\label{factor1}
\langle x_i|K|x_j\rangle&\approx&\left\langle\frac{1}{(2\pi n\beta)^{\frac{1}{2}}}
\sum\limits_{n=1}^{\infty}e^{-\beta n V\left(\frac{x_i+x_j}{2}\right)}
\left(V\left(\frac{x_i+x_j}{2}\right)\right.\right.\nonumber\\&&
\left. \times e^{i\frac{\eta_1(x_i-x_j)}{\sqrt{n\beta}}}\right.\nonumber\\&&
\left.+\frac{1}{2n\beta}\cos\left(\frac{\sqrt{\eta_2^2+\eta_3^2+\eta_4^2}x_i}
{\sqrt{n\beta}}\right)\right.\nonumber\\&&
\left.\cos\left(\frac{\sqrt{\eta_2^2+\eta_3^2+\eta_4^2}x_j}
{\sqrt{n\beta}}\right)\right.\nonumber\\&&
\left.+\frac{1}{2n\beta}\sin\left(\frac{\sqrt{\eta_2^2+\eta_3^2+\eta_4^2}x_i}
{\sqrt{n\beta}}\right)\right.\nonumber\\&&
\left.\left.
\sin\left(\frac{\sqrt{\eta_2^2+\eta_3^2+\eta_4^2}x_j}{\sqrt{n\beta}}
\right)\right)\right\rangle_{\eta}.
\end{eqnarray}
To proceed further we make use of the observation that the correlation
function is very much concentrated along the diagonal $x=x'$
(see Figs. \ref{b1},\ref{b2}).
Based on this fact, we approximate  $V\left(\frac{x_i+x_j}{2}\right)\approx
\sqrt{V(x_i)V(x_j)}$ and $V\left(\frac{x_i+x_j}{2}\right)\approx
\frac{V(x_i)+V(x_j)}{2}$, respectively, in order to achieve a
factorization of the $x$-dependence on the right-hand-side in equ.
(\ref{factor1}). The final step is the introduction of a family of
independent complex Wiener increments 
$d\kappa_{in}$ (with $i=1,2,3$ and $n=1,2,3,\ldots$), i.e.
$d\kappa_{in} d\kappa^*_{jk}=\delta_{ij}\delta_{nk}dt$. With their help
we achieve an explicit expression for the correlated noise
in terms of independent Gaussian ($\eta$) and Wiener ($d\kappa$)
random variables,
\begin{eqnarray}\label{noisewignerweyl}
\langle x_i|\sqrt{K}|d\xi\rangle
&\approx&\sum\limits_{n=1}^{\infty}\frac{e^{-\beta
    n\frac{V(x_i)}{2}}}{(2n\beta\pi)^{\frac{1}{4}}}\left(\sqrt{V(x_i)}
d\kappa_{1n}e^{-i\frac{x_i \eta_1}{\sqrt{n\beta}}}\right.\nonumber\\&&
\left.+\frac{1}{\sqrt{2n\beta}}\right.\nonumber\\&&
\left.\times\left(d\kappa_{2n}\cos\left(\frac{x_i}{\sqrt{n\beta}}
\sqrt{\eta_2^2+\eta_3^2+\eta_4^2}\right)\right.\right.\nonumber\\&&
\left.\left.+d\kappa_{3n}\sin\left(\frac{x_i}{\sqrt{n\beta}}
\sqrt{\eta_2^2+\eta_3^2+\eta_4^2}\right)\right)\right).\nonumber\\
\end{eqnarray}
The good agreement of the correlation function of the noise in Wigner-Weyl 
approximation from (\ref{noisewignerweyl}) and the exact correlation 
function can be seen in Fig. \ref{b2}, where we show on the left an average over numerical
realisations from (\ref{noisewignerweyl}) and on the right the exact correlation function
$\langle x_i|K|x_j\rangle$ for an isotropic harmonic oscillator at $kT=3\hbar\omega$. 

To summarize this section, we are able to obtain the non-trivial noise and 
matrix elements of $K$ with reasonable effort. The other terms of the 
equation can be propagated using Fast-Fourier transformation. 
Thus, we are able to propagate the whole stochastic field equation in 
$D$ spatial dimensions. Results of such numerical simulations are presented 
in the following Section \ref{Results}.

\section{\label{Results}Results of numerical simulations}

Our equation is here applied to a 3D Bose gas of fixed particle number
trapped in various 3D potentials and different quantities are calculated. 
The results are obtained in energy or position representation. 
The applicability of our equation is first shown by determining
the Wigner function 
$W(x,p)=\frac{1}{2\pi}\int dy\, 
e^{ipy}\langle\hat{\psi}^{\dagger}(x+\frac{y}{2})\hat{\psi}(x-\frac{y}{2})\rangle$ of a Bose gas. 
The propagation itself is performed in 3D position space, the
figures show the Wigner function integrated over the remaining four phase
space coordinates.
As already explained in Section \ref{ni} we use a Wigner-Weyl 
approximation for the simulation of (\ref{quantumequationIto}).

In Fig. \ref{b3} we show
the Wigner function for an ideal gas of $100$ particles for different 
temperatures and for an isotropic 3D harmonic potential (left) and 
the potential $V(x,y,z)=x^4+y^4+z^4$ (right). The left columns in both
figures show a snapshot of a single realization, while the right columns
display an averaged result over 15000 time steps. Clearly, to arrive at
the equilibrium distribution, an average over many runs is required. 
The temperatures are chosen above, at, and below the critical 
temperature $T_c(N)$ (for the finite $N$ particle system). 
For the harmonic potential (left) one sees the expected qualitative properties;
when applied to the potential $V(x,y,z)=x^4+y^4+z^4$ (right) we see that 
our implementation can be easily used for arbitrary trap potentials
with an unknown energy spectrum.

\begin{figure}[h,t]
  \centering
  \fbox{
    \includegraphics[width=8.0cm]{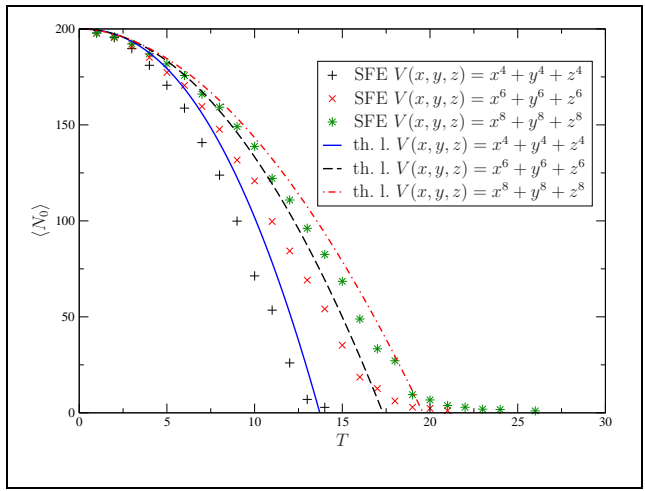}

  }
  \caption{Ground state occupation as a function of temperature from a 
simulation of an ideal 3D Bose gas of 200 particles trapped in the potentials 
$V(x,y,z)=x^4+y^4+z^4$ (plus signs), $V(x,y,z)=x^6+y^6+z^6$ (crosses) and 
$V(x,y,z)=x^8+y^8+z^8$ (stars). We compare with the thermodynamic limit 
(full: $x^4$,
dashed: $x^6$, dashed-dotted:$x^8$).
}\label{b4}
 \end{figure}

\begin{figure}[t]
  \centering
  \fbox{
    \includegraphics[width=8.5cm]{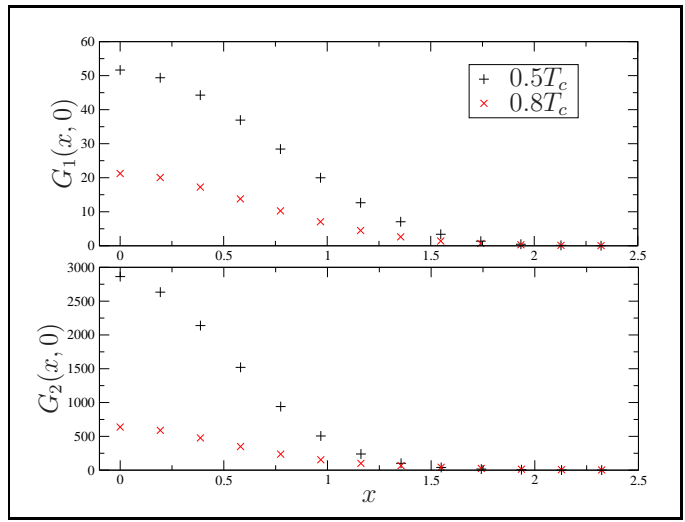}

  }
  \caption{First and second order spatial correlations $G_1(x,0)$ (top) and
  $G_2(x,0)$ (bottom) obtained with our stochastic field equation for a 
Bose gas trapped in the potential $V(x,y,z)=x^4+y^4+z^4$ for 
  different temperatures ($0.5 T_c$: crosses, $0.8 T_c$: stars).}
\label{b5}
 \end{figure}

\begin{figure}[h,t]
  \centering
  \fbox{
    \includegraphics[width=8.5cm]{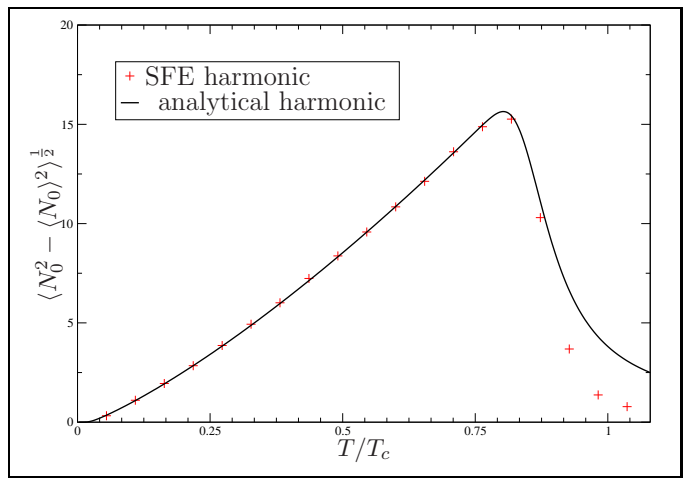}

  }

  \caption{Simulation (plus signs) of the variance of the ground state 
occupation for a 3D ideal Bose gas in an isotropic harmonic potential 
compared to results from \cite{Koc00} (full line).}
\label{b6}
 \end{figure}

We next show that we also obtain good results quantitatively. First,
we focus on the ground state occupation of the ideal gas. 
In \cite{Hel09} we showed the good agreement of our values with other 
canonical results for a box and a harmonic potential. In Fig. \ref{b4} 
results for the condensed fraction in potentials with higher powers 
than the harmonic case are shown. We use the 3D potentials 
$V(x,y,z)=x^4+y^4+z^4$ (plus signs), $V(x,y,z)=x^6+y^6+z^6$ (crosses) and 
$V(x,y,z)=x^8+y^8+z^8$ (stars) also to illustrate the flexibility of
the numerical implementation of our equation.
We compare with the results obtained in the thermodynamic limit 
(full line, dashed, dashed dotted); clearly, finite size effects
are visible.

Spatial correlation functions are quantities which are often measured in 
experiments. In \cite{Hel09} we investigated second order correlation 
functions and their remarkable differences between a canonical and a
grand canonical description. We also showed the good agreement of our 
values with corrected grand canonical results from \cite{Nar99}. Here 
we want to go beyond this well known results and show correlation 
functions for the potential $V(x,y,z)=x^4+y^4+z^4$. In Fig. \ref{b5} 
we present calculations of first 
$G_1(x,0)=\langle \hat{\psi}^{\dagger}(x)\hat{\psi}(0)\rangle$ and 
second order $G_2(x,0)=\langle \hat{\psi}^{\dagger}(x)
\hat{\psi}^{\dagger}(0)\hat{\psi}(x)\hat{\psi}(0)\rangle$ correlation 
functions as they depend on the $x$-coordinate for a Bose gas 
of $200$ particles for temperatures below $T_c$ (plus signs, crosses).

The differences of the grand canonical and the canonical ensemble 
become very obvious when considering the fluctuations of the ground 
state occupation as a function of temperature. As already mentioned 
in the introduction, in a grand canonical description the variance 
for temperature near zero would be of the order of the particle number 
$N$. In Fig. \ref{b6} we display the ground state number fluctuations 
for a Bose gas of 200 particles in an isotropic harmonic potential 
in the canonical ensemble that clearly tend to zero as temperature lowers. 
We compare our calculation with the analytical canonical
result of \cite{Koc00} which is based on the known 
eigenenergies for the harmonic potential.

So far, for the determination of the graphs in Figs. \ref{b3}-\ref{b5},
is was sufficient to simulate with a single norm as elaborated upon
in Sect. \ref{norm}. Now, we find that the number fluctuations
are very sensitive with respect to slight errors 
of first $\langle N_0\rangle$ or second order $ \langle N^2_0 \rangle$ 
expectation values. Therefore, for this application, in order to
reach the precision shown in Fig. \ref{b6}, it is necessary to
simulate with a distribution of norms. We use the simplified
Poisson-type distribution from (\ref{dis-approx}), which explains 
why near the critical temperature deviations from the exact result occur. 

One can summarize our simulations by pointing out that with our
stochastic field equation (\ref{quantumequation}) in combination
with the Wigner-Weyl approximation for the noise generation, one
gets good numerical results for many different temperature dependent 
quantities in arbitrary potentials, also embracing the fixed particle 
number in these systems.

\section{Conclusion and Outlook}

In this paper we have discussed a novel stochastic field equation for a 
Bose gas with a finite and fixed particle number: our theory is based on 
the canonical ensemble referring to the conditions found in actual experiments
with atomic gases in traps. The equation is exact for ideal gases. A
generalization including atomic interactions appears possible on the
basis of the inclusion of a mean-field interaction term in the potential.

Our approach to a numerical solution of the equation is explained in great detail. We focus 
on the implementation in position space, in which it is not necessary to 
know the single-particle eigenenergies of the system. By changing the specification of the 
external potential in the numerical code, we can easily determine equilibrium
properties of the gas for many trap geometries. We can calculate correlation 
functions of arbitrary order and in principle obtain information about the
full thermal canonical state. We show that it is possible to achieve 
results with satisfying precision for 
many different quantities of the ideal gas like first and second order 
correlation functions, ground state occupations and the variance of the 
ground state occupation.

\section{Acknowledgement}
We are grateful for inspiring discussions with Markus Oberthaler and 
Thimo Grotz. Sigmund Heller acknowledges support from the International 
Max Planck Research School for Dynamical Processes in Atoms, Molecules 
and Solids, Dresden. 

\begin{appendix}

\section{Calculation of the norm distribution 
${\tilde{P}}({\cal N})$ \label{A1}}

In this appendix we derive expression (\ref{pdistribution}) for the 
distribution of the norm,
\begin{eqnarray}
{\tilde{P}}({\cal N})
=\frac{\int d\mu\{z\}\delta({\cal N}-\sum\limits_k|z_k|^2)W_{N-1}(\{z\})}{\int d\mu\{z\}W_{N}(\{z\})}
\label{weightterm_app},    
\end{eqnarray}
with
\begin{eqnarray}
W_{N-1}\{z\}=\frac{1}{(N-1)!}\left(\sum\limits_{i}|z_i|^2\right)^{N-1}e^{-\sum\limits_ie^{\beta \epsilon_i}|z_i|^2}.\nonumber\\
\end{eqnarray}
which we use in Section \ref{norm}.
Suppose the number of considered eigenstates equals $M$, which is
arbitrary. We perform the transformation of variables: 
$|z_i|^2\rightarrow x_i$ and find
\begin{eqnarray}
\label{P-FM}
\tilde{P}({\cal N})&=&\frac{2}{C}\int\limits_{0}^{\infty}d^{(M-1)}x
\int\limits_{0}^{\infty}dx_0\delta(x_0-({\cal N}-\sum\limits_ix_i))
\nonumber\\&&
  \frac{1}{(N-1)!}\left(\sum_ix_i\right)^{(N-1)}
e^{-\sum\limits_ie^{\beta \epsilon_i}x_i}\nonumber\\&=&
\frac{2}{C(N-1)!}{\cal N}^{(N-1)}e^{-e^{\beta \epsilon_0}{\cal N}}
\nonumber\\&&
\times\underbrace{\int\limits_{0<\sum\limits_{i=1}^{M-1}x_i<{\cal N}}d^{M-1}x
  e^{-\sum\limits_{i=1}^{M-1}(
e^{\beta \epsilon_i}-e^{\beta \epsilon_0})x_i}}_{:=F_{M-1}}
\end{eqnarray}
with $C=\int d\mu\{z\}W_N(\{z\})$. First, we calculate the integral $F_M$.
\begin{eqnarray}
F_M&=&\int\limits_0^{{\cal N}}dx_1
\int\limits_{0<\sum\limits_{i=2}^{M-1}x_i<{\cal N}-x_1}
dx_2...\int\limits_{0<x_M<{\cal N}-\sum\limits_{i=1}^{M-1}x_i}
dx_M\nonumber\\&&
\times e^{-\sum\limits_{i=1}^{M}(e^{\beta \epsilon_i}-e^{\beta \epsilon_0})x_i}.
 \end{eqnarray}
Now we substitute $\tilde{x}_1=x_1$, $\tilde{x}_2=x_2+x_1$, 
$\tilde{x}_3=x_3+\tilde{x}_2$ etc. and rewrite the integral as
\begin{eqnarray}
\label{F_M}
F_M&=&\int\limits_{0}^{{\cal N}}d\tilde{x}_1
e^{-(e^{\beta\epsilon_1}-e^{\beta\epsilon_2})\tilde{x}_1}
\int\limits_{\tilde{x}_1}^{{\cal N}}d\tilde{x}_2
e^{-(e^{\beta\epsilon_2}-e^{\beta\epsilon_3})\tilde{x}_2}...\nonumber\\&&
\int\limits_{\tilde{x}_{M-1}}^{{\cal N}}d\tilde{x}_{M}\,
e^{-(e^{\beta\epsilon_M}-e^{\beta\epsilon_0})\tilde{x}_M}.
\end{eqnarray}
By considering a derivative with respect to ${\cal N}$ of equation 
(\ref{F_M}) it can be seen that $F_M$ satisfies the differential equation  
\begin{eqnarray}
\label{differentialequation}
\frac{\partial F_M({\cal N})}{\partial {\cal N}}=e^{-(e^{\beta\epsilon_M}-e^{\beta\epsilon_0}){\cal N}}F_{M-1}({\cal N}).
\end{eqnarray}
We succeeded to find a solution with the ansatz
\begin{eqnarray}
F_M({\cal N})&=&\prod\limits_{k=1}^{M}
\frac{1}{e^{\beta \epsilon_k}-e^{\beta \epsilon_0}}\nonumber\\
&+&\sum\limits_{k=1}^{M}
\frac{g_{k}^{M}}{-e^{\beta \epsilon_k}+e^{\beta \epsilon_0}}
e^{-(e^{\beta \epsilon_k}-e^{\beta \epsilon_0}){\cal N}}\nonumber\\
\end{eqnarray}
with $g_M^{M}=\prod\limits_{k=1}^{M-1}
\frac{1}{e^{\beta \epsilon_k}-e^{\beta \epsilon_M}}$
and $g_k^M=\frac{g_k^{M-1}}{-e^{\beta \epsilon_k}+e^{\beta \epsilon_M}}$. 
In this way we can obtain all coefficients $g_k^M$ iteratively. 
With a bit of analysis we find
\begin{eqnarray}
\label{FMresult}
F_M=\sum\limits_{k=0}^{M}\prod\limits_{\begin{array}{c}
\scriptstyle l=0\scriptstyle \\ \scriptstyle l\neq k
\end{array}}^{M}
\frac{1}{(e^{\beta \epsilon_l}-e^{\beta \epsilon_k})}
e^{-(e^{\beta \epsilon_k}-e^{\beta \epsilon_0}){\cal N}}.
\end{eqnarray}
Finally, we plug expression
(\ref{FMresult}) into equation (\ref{P-FM}) and use 
$C=\int d\mu\{z\}W_N(\{z\})=
\int d{\cal N}\int d\mu\{z\}\delta({\cal N}-\sum\limits_k|z_k|^2)W_N(\{z\})$ 
to write the norm distribution in the form 
\begin{eqnarray}
\tilde{P}({\cal N})=\frac{1}{(N-1)!}
\frac{{\cal N}^{(N-1)}\sum\limits_{k}c_ke^{-e^{\beta \epsilon_k} {\cal N}}}
{\sum\limits_kc_ke^{-\beta \epsilon_k(N+1)}}.\nonumber\\ 
\end{eqnarray}
with $c_k=\prod\limits_{\begin{array}{c}\scriptstyle l=0\scriptstyle \\ \scriptstyle l\neq k
\end{array}}\frac{1}{(e^{\beta \epsilon_l}-e^{\beta \epsilon_k})}$,
as it is written in equation (\ref{pdistribution}) of Section \ref{norm}.
\end{appendix}

\end{document}